\documentclass[conference,a4paper]{IEEEtran}
    \newtheorem{theorem}{Theorem}
    \newtheorem{lemma}[theorem]{Lemma}

\usepackage[cmex10]{amsmath}
\usepackage{mathtools, amssymb, mleftright, booktabs}
    \def\ee{\varepsilon}
    \def\NN{{\mathbf N}}
    \def\ZZ{{\mathbf Z}}
    \def\Wp{W_\mathrm p}
    \def\Wg{W_\mathrm g}
    \DeclareMathOperator\Prob{Prob}
    \DeclareMathOperator\Hash{Hash}
    \DeclareMathOperator\Length{Length}

\usepackage{tikz-cd}
    \tikzset{every picture/.style={cap=round, join=round}}

\usepackage[noadjust]{cite}
\usepackage{hyperref}
    \hypersetup{
        colorlinks, allcolors=green!20!blue!75!black,
        pdfsubject={Information Theory (cs.IT)},
        pdfkeywords={
            group testing,
            binary splitting,
            fast decoder,
            capacity achieving,
            branching process,
            Catalan numbers
        },
    }

\begin{document}

\advance\baselineskip 0pt plus.1pt minus.05pt
\advance\lineskip 0pt plus.1pt minus.05pt
\advance\parskip 0pt plus.2pt minus.1pt

\title{Quickly-Decodable Group Testing with Fewer Tests:
    Price--Scarlett and Cheraghchi--Nakos's
    Nonadaptive Splitting with Explicit Scalars}

\author{%
    \IEEEauthorblockN{Hsin-Po Wang}%
    \IEEEauthorblockA{%
        University of California, Berkeley, CA\\
        simple@berkeley.edu%
    }
    \and
    \IEEEauthorblockN{Ryan Gabrys}%
    \IEEEauthorblockA{%
        University of California San Diego, CA\\
        Naval Information Warfare Center, CA\\
        rgabrys@ucsd.edu%
    }
    \and
    \IEEEauthorblockN{Venkatesan Guruswami}%
    \IEEEauthorblockA{%
        University of California, Berkeley, CA\\
        venkatg@berkeley.edu%
    }
}

\maketitle

\begin{abstract}\boldmath
    We modify Cheraghchi--Nakos \cite{ChN20} and Price--Scarlett's
    \cite{PrS20} fast binary splitting approach to nonadaptive group
    testing.  We show that, to identify a
    uniformly random subset of \(k\) infected persons among a population
    of \(n\), it takes only \(\ln(2 - 4\varepsilon) ^{-2} k \ln n\)
    tests and decoding complexity \(O(\varepsilon^{-2} k \ln n)\), for
    any small \(\varepsilon > 0\), with vanishing error probability.  In
    works prior to ours, only two types of group testing schemes exist.
    Those that use \(\ln(2)^{-2} k \ln n\) or fewer tests require
    linear-in-\(n\) complexity, sometimes even polynomial in \(n\);
    those that enjoy sub-\(n\) complexity employ \(O(k \ln n)\) tests,
    where the big-\(O\) scalar is implicit, presumably greater than
    \(\ln(2)^{-2}\).  We almost achieve the best of both worlds, namely,
    the almost-\(\ln(2)^{-2}\) scalar and the sub-\(n\) decoding
    complexity.  How much further one can reduce the scalar
    \(\ln(2)^{-2}\) remains an open problem.
\end{abstract}

\section{Introduction}

    % why we study GT
    In the past few years, group testing (GT) has been used to fight
    against Covid \cite{AlE22}.  Before that, it was used in genotyping,
    finding heavy-hitters, and scheduling multiaccess channels, and is
    also related to compressed sensing and machine learning~\cite{DuH93,
    AJS19}.

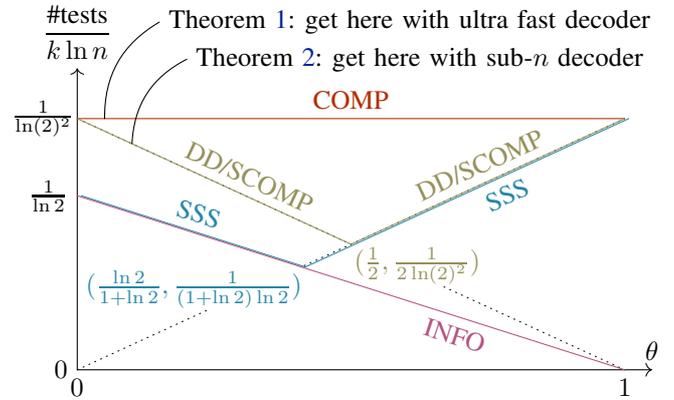
\begin{figure}
    \begin{tikzpicture} [x=72mm, y=16mm]
        \draw [->] (0, 0) -- (0, 2.5)
            node [above] {$\dfrac{\text{\#tests}}{k\ln n}$};
        \draw [->] (0, 0) -- (1.05, 0) node [above] {$\theta$};
        \path
            (0, {1/ln(2)^2}) coordinate (NW)
            node [left, inner sep=1pt] {$\frac{1}{\ln(2)^2}$}
            (1, {1/ln(2)^2}) coordinate (NE)
            (0, {1/ln(2)}) coordinate(W) node [left]  {$\frac{1}{\ln2}$}
            (0, 0) coordinate (SW) node [left] {$0$} node [below] {$0$}
            (1, 0) coordinate (SE) node [below] {$1$};
        \draw [dotted] (NW) --  (SE) (SW) -- (NE);
        \draw [orange!50!red!70!black] (NW) -- node [above] {COMP} (NE);
        \draw [yellow!50!black]
            (NW) -- node [pos=.6, auto, sloped] {DD/SCOMP}
            (1/2, {1/(2*ln(2)^2)}) node
            [below right, inner sep=1pt, fill=white]
            {$(\frac12, \frac{1}{2\ln(2)^2})$}
            -- node [auto, sloped] {DD/SCOMP} (NE);
        \draw [cyan!60!black, transform canvas={xshift=1.5pt}]
            (W) -- node [auto, sloped] {SSS}
            ({ln(2)/(1+ln(2))}, {1/(ln(2)*(1+ln(2)))})
            node [below left, inner sep=1pt, fill=white]
            {$(\frac{\ln2}{1+\ln2}, \frac{1}{(1+\ln2)\ln2})$}
            -- node [pos=.6, auto, ', sloped] {SSS} (NE);
        \draw [magenta!70!black]
            (W) -- node [pos=.7, auto, ', sloped] {INFO} (SE);
        \draw (.05, {1/ln(2)^2}) to [bend left=30] +(.1, .8) node [right]
            {Theorem~\ref{thm:COMP}: get here with ultra fast decoder}
            (.1, {0.9/ln(2)^2}) to [bend left=30] +(.1, .7) node [right]
            {Theorem~\ref{thm:DD}: get here with sub-$n$ decoder};
    \end{tikzpicture}
    \caption{
        The code rates of GT strategies.
        Horizontal axis is $\theta = \log_n k$.
        COMP's rate is
        pinpointed in \cite[Theorem~2]{JAS19}.
        DD's rate is
        lower bounded in \cite[Theorem~3]{JAS19} and
        upper bounded in \cite[Theorem~1.2]{CGH20}.
        SCOMP's rate is
        upper bounded in \cite[Theorem~1.2]{CGH20} and
        lower bounded by an argument around \cite[formula~(1)]{CGH20}.
        SSS stands for smallest satisfying set,
        the peak of nonadaptive (one-stage) GT;
        its rate is
        upper bounded in \cite[Theorem~4]{JAS19} and 
        lower bounded in \cite[Theorem~1.1]{CGH20}
        (Cf. \cite[Corollary~2.1]{BSP22} and \cite[Theorem~4]{FlM21}).
        A polynomial-time design, SPIV,
        achieves the rate of SSS \cite[Theorem~1.2]{CGH21}.
        INFO is the naive counting bound.
        INFO is attained by a two-stage GT, the least possible number
        of stages for adaptive GT \cite[Theorem~1.3]{CGH21}.
        Similar plots are found in
        \cite[Figure~1]{CGH21} and \cite[Figure~2]{CGH20}.
        Up-side-down plots are found in \cite[Figure~2.1]{AJS19},
        \cite[Fig.~1]{JAS19}, \cite[Figure~3]{ScC17},
        \cite[Fig.~1]{Ald17}, and \cite[Fig.~2]{ABJ14}.
    }                                               \label{fig:capacity}
\end{figure}

    % GT and IT
    Recently, information theory has shed new light on GT \cite{AtS09,
    ScC17, AJS19, CGH20}.  For one, coding techniques such as list
    decoding \cite{INR10}, Kautz--Singleton \cite{IKW19}, SC-LDPC
    \cite{CGH21}, and error-locating codes \cite{LCP19, BCS21} enter GT.
    For another, the notion of \emph{capacities} is introduced
    \cite{BJA13, Ald17}.  For instance, the standard argument is that
    since there are $\binom{n}{\leq k}$ possible sets of infected
    persons, and since the test outcome is binary, there should be at
    least $\log_2 \binom{n}{\leq k}$ tests, as represented by the INFO
    line in Fig.~\ref{fig:capacity}.

    % incoherent capacity issue
    To meet INFO, tests should be positive with probability $p \approx
    1/2$.  However, when $\ln k > 0.41 \ln n$, half of tests being
    positive is so ``polluted'' that some healthy person will have all
    associated tests positive; it is then impossible to tell if this
    person is healthy.  So nonadaptive GT does not always stays on the
    INFO line, but on the SSS curve in Fig.~\ref{fig:capacity}.
    Sometimes, a decoder with less complexity---such as COMP, DD, and
    SCOMP \cite[Section~2.1]{AJS19} \cite[Section~5]{wiki-gt}---is
    favored over the omnipotent decoder SSS, but they do not share the
    same code rate, as illustrated by Fig.~\ref{fig:capacity}.

\begin{table}
    {
    \centering
    \caption{
        Some low-complexity GT strategies, extending
        \cite[Table~1]{PrS20}.
    }                                                   \label{tab:fast}
    \def\arraystretch{1.2}
    \tabcolsep3pt
    \begin{tabular}{ccc}
        \toprule
        papers
        & number of tests
        & decoding complexity                                         \\
        \midrule
        SAFFRON \cite{LCP19}
        & $O(k \ln(k) \ln n)$
        & $O(k \ln k)$                                                \\
        GROTESQUE \cite{CJB17}
        & $O(k \ln(k) \ln n)$
        & $O(k \ln(k) \ln n)$                                         \\
        via Kautz--Singleton \cite{IKW19}
        & $O(k \ln(n) \ln \frac{\ln n}{\ln k})$
        & $O(k^3 \ln(n) \ln \frac{\ln n}{\ln k})$                     \\
        bit-mixing coding \cite{BCS21}
        & $O(k \ln n)$
        & $O(k^2 \ln(k) \ln n)$                                       \\
        Price--Scarlett \cite{PrS20}$^\triangleright$
        & $O(k \ln n)$
        & $O(k \ln n)$                                                \\
        Cheraghchi--Nakos \cite{ChN20}$^\triangleright$
        & $O(k \ln n)$
        & $O(k \ln n)$                                                \\
        \raisebox{-1pt}{Theorem~\ref{thm:COMP}}
        & $\frac{k}{\ln(2 - 4\ee)^2} \ln n$ 
        & $O(\frac{k}{\ee^2} \ln n)$                                  \\
        Theorem~\ref{thm:DD}\rlap{$^\star$}
        & $\frac{k}{\ln(2 - 4\ee)^2} \ln \frac{2n}{\ee k}$ 
        & $O(\frac{k^2}{\ee^2} \ln n)$                                \\
        SPIV \cite{CGH21}
        & $\frac{k}{\ln 2} \max(\ln\frac nk, \log_2 k)$
        & polynomial in $n$                                           \\
        \bottomrule
    \end{tabular}
    }

    \smallskip \tiny
    $^\triangleright$Cheraghchi--Nakos \cite[Theorem~20]{ChN20}
    \cite[Theorem~3.10]{ChN20x} and Price--Scarlett \cite{PrS20}
    are concurrent.

    $^\star$Theorem~\ref{thm:DD} assumes $\theta = \log_n k < 1/2$.
    \vskip-6pt
\end{table}

    % fast decoder
    Whereas the works mentioned in Fig.~\ref{fig:capacity} have
    complexities $O(n)$ or higher, there are GT schemes with
    \emph{sub-$n$ decoders}, such as those in Table~\ref{tab:fast}.  To
    reduce complexity, more tests are used.  So for instance, the
    big-$O$ factor of bit-mixing coding \cite{BCS21} is $\geq 18$; and
    for Price--Scarlett \cite{PrS20}, it is $16$.

    % we want both
    In this work, we keep the cake and eat it too.  We take
    Cheraghchi--Nakos \cite{ChN20} and Price--Scarlett's \cite{PrS20}
    sub-$n$ scheme (hereafter PCNS) and redo the
    analysis using techniques such as Möbius transformations and
    generating functions to minimize the big-$O$ factor.  We manage to
    match the factor of COMP.  We also provide a variant with a slightly
    higher complexity whose number of tests matches that of DD.

    % concurrent works
    It is worth noting that Cheraghchi--Nakos \cite[Theorem~20]{ChN20},
    \cite[Theorem~3.10]{ChN20x} concurrently obtained the same result as
    PCNS.  The proof ideas are similar and we thank the reviewer for the
    reminder.

    This paper is organized as follows.
    Section~\ref{sec:state} states the problem and main results.
    Section~\ref{sec:PCNS} summarizes PCNS.
    Section~\ref{sec:COMP} proves Theorem~\ref{thm:COMP}.
    Section~\ref{sec:DD} proves Theorem~\ref{thm:DD}.

\section{Problem Statement and Main Results}           \label{sec:state}

    % define GT
    Let $k$, $m$, and $n$ be positive integers.  Let $x \in \{0, 1\}^n$
    be an unknown vector with Hamming weight $k$ or less.  We will
    consider the sparse regime $k = n^\theta$, where $\theta \in (0, 1)$
    is called the \emph{sparsity parameter}.  We assume that each of $n$
    choose $k$ vectors is selected by Mother Nature with equal
    probability.  Let $G \in \{0, 1\}^{m\times n}$ be the
    \emph{measurement matrix}.  Let $y \coloneqq Gx \in \{0, 1\}^m$ be
    the \emph{test outcomes}, wherein the scalar multiplication and
    scalar addition are replaced by logical-and and logical-or,
    respectively:
    \[ y_i \coloneqq \bigvee_j (G_{ij} \wedge x_j). \]
    Our goal is to design $G$ and observe $y$, hoping that we can
    recover $x$.  To be more precise, we want a decoder that maximizes
    $\Prob \{ \mathrm{Decode} (G, y) = x \}$ while minimizing the number
    of tests $m$ and the complexity of the decoder.  Our main results
    are as follows.

    \begin{theorem} [Attaining COMP's code rate]        \label{thm:COMP}
        Fix a $\theta \in (0, 1)$ and a sufficiently small $\ee > 0$.
        For $n$ large enough, there exists a measurement matrix that
        (a) performs $m = \ln(2 - 4\ee)^{-2} \* k \* \ln n$ tests,
        (b) pairs with a decoder with $O(\ee^{-2} \* k \* \ln n)$
            operations,
        (c) produces no false negatives, and
        (d) produces one or more false positives with probability at
            most $2 / \ee k^{2\ee} + e^{-k} + 2n^{-k}$.
    \end{theorem}

    In the previous theorem, we allow false positives.  In the next
    theorem, we allow false negatives.

    \begin{theorem} [Attaining DD's code rate]            \label{thm:DD}
        Fix a $\theta \in (0, 1/2)$ and a sufficiently small $\ee > 0$.
        For $n$ large enough, there is a measurement matrix that
        (a) performs $m = \ln(2 - 4\ee)^{-2} \* k \* \ln(2n / \ee k)$
            tests,
        (b) pairs with a decoder with $O(\ee^{-2} \* k^2 \* \ln n)$
            operations,
        (c) produces no false-positive, and
        (d) produces one or more false negatives with vanishing
            probability.
    \end{theorem}
    
    Note that the $k^2$ term in (b) is sub-$n$ since we assume $\theta <
    1/2$.  If $\theta > 1/2$, our GT scheme degenerates to DD.

    % Taxonomy
    According to Aldridge, Johnson, and Scarlett's classification
    \cite[Chapter~1]{AJS19} of the space of GT, this work
    studies \textbf{nonadaptive} tests,
    allows \textbf{small error probability}
    (in the \textbf{for-each} setting),
    aims for \textbf{exact recovery},
    requires \textbf{noiseless testing},
    receives \textbf{binary outcome},
    models with \textbf{combinatorial prior},
    assumes \textbf{known number of defectives},
    contributes to the \textbf{sparse regime},
    and uses a \textbf{constant tests-per-item design}.

\section{PCNS's Strategy}                               \label{sec:PCNS}

    % introduce prefix
    Suppose that $k$ and $n$, numbers of infected and total population,
    are powers of $2$.  Each of $n$ persons is labeled by a binary
    string of length $\log_2 n$.  By a \emph{prefix} of length $\ell$,
    where $1 \leq \ell \leq \log_2 n$, we mean the first $\ell$ bits of
    a binary string.  A prefix $P \in \{0, 1\}^\ell$ of length $\ell$
    \emph{includes} $n/2^{\ell}$ persons---those whose labels begin with
    $P$.  We say a prefix is \emph{innocent} if everyone it includes is
    healthy, \emph{suspicious} if we cannot prove it innocent yet.

\subsection{Phase I of PCNS: grow-and-prune}

    % Phase I testing schedule
    PCNS has two phases.  Phase I consists of $16k \log_2(n/k)$ tests
    indexed by $\{\log_2(k) + 1,\allowbreak \dotsc, \log_2 n\} \times
    \{0, \dotsc, 16k - 1\}$.  For each length $\ell = \log_2(k) + 1,
    \dotsc, \log_2 n$, every possible prefix $P \in \{0, 1\}^\ell$ is
    fed into a hash function $\Hash\colon \bigcup_{\ell=0}^\infty
    \ZZ^\ell \to \NN$.  Persons having prefix $P$ are assigned to the
    test labeled $(\ell, \Hash(P) \bmod 16k)$.  (See
    \cite[Section~3]{PrS20} for a discussion on the quality of the hash
    function.  For this paper, we assume that the hash function
    generates a true random number and remember this output for this
    input.)

    % Phase I decoding algorithm, initialization
    To decode Phase I, dubbed \emph{grow-and-prune}, we maintain two
    watch lists, $\Wp$ and $\Wg$, of suspicious prefixes.  We initialize
    $\Wp$ with all prefixes of length $\log_2 k$.

    % Phase I decoding algorithm, iteration core
    For each prefix $P \in \Wp$, put $P0$ and $P1$ on the other watch
    list $\Wg$, and empty $\Wp$ afterward.  This is the \emph{growing}
    part of Phase I.  For the \emph{pruning} part of Phase I, for every
    prefix $Q \in \Wg$, check the outcome of the test $(\Length(Q),
    \Hash(Q) \bmod 16k)$.  A positive outcome means that persons having
    prefix $Q$ are suspicious, so we move $Q$ back to the watch list
    $\Wp$.  A negative outcome means that persons having prefix $Q$ are
    all healthy, so we discard $Q$ from $\Wg$.

    % Phase I decoding algorithm, iteration boundaries
    Every time we repeat the previous paragraph once, the length of the
    prefixes in $\Wp$ increases by $1$.  Therefore, $\Wp$ will end up
    with strings of length $\log_2 n$, each indicating a suspicious person.

\subsection{Phase II of PCNS: leaf-trimming}

    % Phase II's input
    Phase II of PCNS continues with the watch list $\Wp$ and
    double-checks persons on it to remove as many uninfected persons as
    possible.

    % Phase II testing schedule
    Now prepare $16k \log_2 k$ tests indexed by $\{1, \dotsc, \log_2 k\}
    \times \{0, \dotsc, 16k - 1\}$.  For all $\ell = 1, \dotsc, \log_2
    k$ and all labels $S \in \{0, 1\}^{\log_2 n}$, assign the person
    with label $S$ to the test labeled $(\ell, \Hash(\ell, S) \bmod
    16k)$ (note that the hash function takes a length-$(1 + \log_2 n)$
    vector as input).

    % Phase II decoding algorithm
    To decode Phase II, consider a person $S \in \Wp$ and check the
    outcome of $(\ell, \Hash(\ell, S) \bmod 16k)$ for all $\ell = 1,
    \dotsc, \log_2 k$.  If any of these tests is negative, remove $S$.
    Finally, we declare that persons still on $\Wp$ as infected.

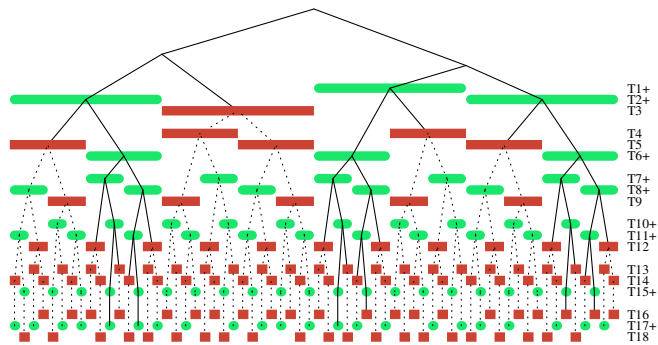
\begin{figure}
    \centering
    \makeatletter
    \def\pgfkeysgsetvalue#1#2{%
        \pgfkeys@temptoks{#2}%
        \expandafter\xdef\csname pgfk@#1\endcsname
            {\the\pgfkeys@temptoks}%
    }
    \pgfkeys{/handlers/.ginitial/.code={
        \pgfkeysgsetvalue{\pgfkeyscurrentpath}{#1}
    }}
    \pgfmathdeclarefunction{Hash}{2}{%
        \def\L{#1}%
        \def\P{floor(#2)}%
        \ifnum #1 > 6\relax%
            \def\P{floor(#2 * 3 * .618)}%
        \fi%
        \pgfmathparse{mod(\P + 1, 3) - 1}%
    }
    \makeatother
    \tikzset{
        positive/.style={
            green!60!cyan!80!gray, line cap=round, blend mode=darken},
        negative/.style={red!80!orange!60!gray, line cap=rect}
    }
    \begin{tikzpicture} [x=5mm, y=6mm]
        \def\infectedA{13}
        \def\infectedB{34}
        \foreach \L in {0, ..., 7}{
            \pgfmathsetmacro\twotoL{2^min(\L, 6)}
            \pgfmathsetmacro\twotoLminusone{\twotoL - 1}
            \pgfmathtruncatemacro\posiA{Hash(\L, \infectedA*\twotoL/64)}
            \pgfmathtruncatemacro\posiB{Hash(\L, \infectedB*\twotoL/64)}
            \pgfkeys{/outcome/\posiA/.initial={+}}
            \pgfkeys{/outcome/\posiB/.initial={+}}
            \ifnum \L > 1
                \foreach \Tmod in {-1, 0, 1}{
                    \pgfkeys{/outcome/\Tmod/.get=\outcome}
                    \pgfmathtruncatemacro\Tid{\L*3 + \Tmod - 4}
                    \node at (16, -\L - \Tmod/4) [right]
                    {\tiny T\Tid\outcome};
                }
            \fi
            \foreach \P in {0, ..., \twotoLminusone}{
                \pgfmathtruncatemacro\Tmod{Hash(\L, \P)}
                \pgfkeys{/outcome/\Tmod/.get=\outcome}
                \draw ({(16*\P + 8)/\twotoL}, - \L - \Tmod/4)
                    coordinate(L\L/P\P);
                \ifnum \L > 0
                    \pgfmathtruncatemacro\LL{\L - 1}
                    \pgfmathtruncatemacro\PP{\L>6 ? \P : floor(\P/2)}
                    \pgfkeys{/innocent/\LL/\PP/.get=\innocent}
                    \ifx \innocent \relax
                        \draw (L\L/P\P) -- (L\LL/P\PP);
                    \else
                        \draw [dotted] (L\L/P\P) -- (L\LL/P\PP);
                        \pgfkeys{/innocent/\L/\P/.ginitial={innocent}}
                    \fi
                \fi
                \ifnum \L > 1
                    % \PackageWarning{bonas}{\L, \P; \Tmod, \outcome}
                    \ifx \outcome \relax
                        \tikzset{cell col/.style=negative}
                        \pgfkeys{/innocent/\L/\P/.ginitial={innocent}}
                    \else
                        \tikzset{cell col/.style=positive}
                    \fi
                    \draw [line width=1.25mm, cell col]
                        ({(16*\P + 0)/\twotoL + 1/8}, - \L - \Tmod/4) --
                        ({(16*\P + 16)/\twotoL - 1/8}, - \L - \Tmod/4);
                \fi
            }
        }
    \end{tikzpicture}
    \caption{
        The encoding matrix (strips) and the decoding process (tree).
        Each person is represented by a column $1/8$ centimeters ($1/20$
        inches) wide.  T1--T15 is Phase I; T16--T18 is Phase II.  The
        colored strips are places where the encoding matrix has $1$.  A
        green round strip means that the test turns out positive; a red
        rectangular strip means negative.  A person is declared infected
        if it is all green light when traveling from the root to the
        bottom. See also \cite[Figure~1]{PrS20}.
    }                                                   \label{fig:tree}
\end{figure}

    Fig.~\ref{fig:tree} visualizes Phases I and II.

\subsection{Number of tests}

    Phase I spends $16k \log_2(n/k)$ tests and Phase II spends $16k
    \log_2 k$ tests.  In total, $16k \log_2 n$ tests are spent.

\subsection{Two types of error events: WA and TLE}

    % WA
    We want to analyze two error events of PCNS.  The first type is the
    wrong-answer (WA) event: the set of persons we declared is not
    exactly the set of actually infected persons.  This is the same
    event as that at least one uninfected person remains on $\Wp$ when
    Phase II ends.

    % TLE
    The second type is that we perform too many insertion, removal,
    hashing, or outcome-query operations.  Suppose that we can show
    that, with high probability, the number of operations is $O(k \ln
    n)$.  If we then set a timer that forcibly terminates the decoder
    after $O(k \ln n)$ operations, we will introduce the
    time-limit-exceeded (TLE) event which adds a small amount to the
    overall error probability.  But then we get a \emph{fast} decoder
    with a guaranteed $O(k \ln n)$ runtime. 

\subsection{WA analysis: how many remain on the watch list?}
                                                       \label{sec:PS-WA}

    % biased branching process
    In Phase I, since there are $16k$ tests for every $\ell$, and since
    there are $k$ infected persons, the probability that a test contains
    some infected persons is at most $1/16$.  As a result, each prefix
    that includes no infected persons will have a probability of at
    least $15/16$ to be pruned away.

    % control the population of the $\ell$th generation
    To facilitate discussion, let's assume for now that each prefix of
    length $\log_2 k$ includes exactly one infected person.  Define the
    \emph{excess size} of the watch list as $N_\ell \coloneqq |\Wg| - k$
    when $\Wg$ contains length-$\ell$ prefixes.  The excess sizes are
    controlled by the probability-generating functions $F_\ell(q)
    \coloneqq \sum_{j=0}^\infty \Prob \{N_\ell{=}j\} \cdot q^j$ that
    satisfy this recurrence relation:
    \begin{align}
        F_{\log_2 k} (q)
        & \coloneqq 1, \notag\\
        F_{\ell+1} (q)
        & \coloneqq F_\ell \Bigl( \frac{15}{16} + \frac{q^2}{16} \Bigr)
            \cdot q^k. \label{for:q^k}
    \end{align}
    $15/16$ means that with probability $15/16$ a prefix is pruned;
    $q^2/16$ means that with probability $1/16$ a prefix gives birth to
    two; $q^k$ means that the $k$ infected persons will each give birth
    to an innocent prefix.

    % Mean of N_\ell
    To get a grasp on the magnitude of $N_\ell$, especially $N_{\log_2
    n}$, note that the mean of $N_\ell$ is the derivative of $F_\ell$ at
    $q = 1$, and that the derivatives follow this recurrence relation:
    \begin{align*}
        F_{\log_2 k}' (1)
        & = 0, \\
        F_{\ell+1}' (1)
        & = F_\ell'(1) \cdot \frac{2}{16} + k
            = \Bigl( F_\ell'(1) - \frac{8k}{7} \Bigr) \cdot \frac18
            + \frac{8k}{7}.
    \end{align*}
    This means that the mean of $N_\ell$ is attracted to a fixed point
    $\lim_{\ell\to\infty} F_\ell'(1) = 8k/7$.  Cf.\ \cite[(4)]{PrS20}.

    % Iteration of p.g.f.
    The attraction behavior can also be seen on the polynomial level.
    Suppose $f(q) \coloneqq (15 + q^2) / 16$.  Then it is
    straightforward from the recurrence relation that
    \begin{equation}
        F_\ell(q)                                       \label{for:prod}
        = \prod_{i=0}^{\ell-\log_2k} f^{\circ i} (q) ^k,
    \end{equation}
    where $f^{\circ i}$ is the $i$th iteration of $f$, e.g., $f^{\circ
    2} (q) = f(f(q))$.  To study the tail behavior of $N_\ell$, it
    suffices to study the evaluations of $F_\ell$, which amounts to
    studying $f^{\circ i}$.

    % upper bound by Möbius transformation
    Let $g(q) \coloneqq 6/(7 - q) = f(q) + (q - 1) (q - 3)^2 / 16 (7 -
    q)$.  Clearly, $f(q) \leq g(q)$ for $q \in [1, 7]$.  Observe that
    $g$ is a Möbius transformation that maps $[1, 6]$ to $[1, 6]$ and
    fixes $1$ and $6$.  Thus $g^{\circ i}$, regardless of $i$, is again
    a Möbius transformation mapping $[1, 6]$ to itself and fixing the
    end points.  To determine $g^{\circ i}$ completely (as a Möbius
    transformation is determined by three complex parameters), we
    compute the derivatives at $q = 1$:
    \begin{align}
        g'(1)
        & = 1/6, \notag\\
        (g^{\circ i+1})' (1)
        & = (g^{\circ i})' (1) \cdot g'(1).             \label{for:diff}
    \end{align}
    $g^{\circ i \prime}(1) = 6^{-i}$ allows us to write down $g^{\circ
    i}$ explicitly \[ g^{\circ i} (q) = 1 + \frac {5 (q - 1)} {6^i (6 -
    q) + (q - 1)}. \] In particular, $g^{\circ i} (5) = 1 + 20 / (6^i +
    4)$.

\begin{figure}
    \centering
    \pgfmathdeclarefunction{f}{1}{\pgfmathparse{(15 + #1*#1)/16}}
    \pgfmathdeclarefunction{g}{1}{\pgfmathparse{6/(7 - #1)}}
    \begin{tikzpicture}[x=5mm, y=5mm]
        \draw [->] (1, 1) node [left] {$1$} -- (1, 7);
        \draw [->] (1, 1) node [below] {$1$} -- (7, 1);
        \draw (1, 1) -- (7, 7);
        \draw [blue!36!magenta!60!black]
            plot [domain=1:7, samples=120] (\x, {f(\x)})
            node [right] {$f$};
        \def\xx{5.9}
        \foreach \i in {1, ..., 5}{
            \pgfmathsetmacro\yy{f(\xx)}
            \draw [blue!36!magenta!60!black]
                (\xx, \xx) -- (\xx, \yy) -- (\yy, \yy);
            \xdef\xx{\yy}
        }
    \end{tikzpicture}
    \qquad
    \begin{tikzpicture}[x=5mm, y=5mm]
        \draw [->] (1, 1) node [left] {$1$} -- (1, 7);
        \draw [->] (1, 1) node [below] {$1$} -- (7, 1);
        \draw (1, 1) -- (7, 7);
        \draw [blue!36!magenta!60!black]
            plot [domain=1:7, samples=120] (\x, {f(\x)})
            node [right] {$f$};
        \draw [green!64!yellow!40!black]
            plot [domain=1:7, samples=120] (7 - 6/\x, \x)
            node [above] {$g$};
        \def\xx{5.9}
        \foreach \i in {1, ..., 5}{
            \pgfmathsetmacro\yy{g(\xx)}
            \draw [green!64!yellow!40!black]
                (\xx, \xx) -- (\xx, \yy) -- (\yy, \yy);
            \xdef\xx{\yy}
        }
    \end{tikzpicture}
    \caption{
        Left: iteration of $f$.  Right: iteration of $g$.
    }                                                \label{fig:iterate}
\end{figure}
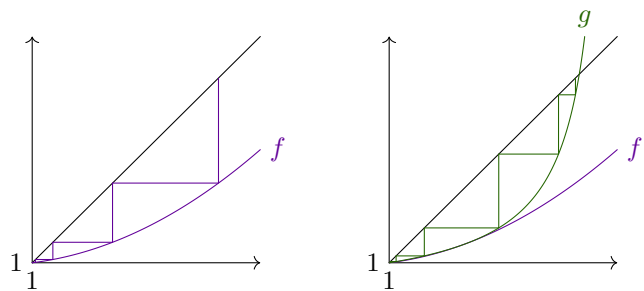

    % apply Chernoff bound
    Back to $f$.  We infer that $f^{\circ i} (5) \leq g^{\circ i} (5)$.
    (See Fig.~\ref{fig:iterate} for an illustration.)
    Hence,
    \[
        F_\ell(5)
        \leq \prod_{i=0}^{\ell-\log_2k} g^{\circ i} (5) ^k
        \leq \exp \Bigl( \sum_{i=0}^\infty \frac{20}{6^i + 4} \Bigr) ^k
        \leq \exp(7k).
    \]
    Recall that $F_\ell$ is the probability-generating function of
    $N_\ell$.  Apply the Chernoff bound:
    \[
        \Prob \{N_\ell{\geq}5k\}
        \leq \frac {F_\ell(5)}{5^{5k}}
        \leq e^{-k}.
    \]
    This inequality holds for all $\ell$, in particular $\ell = \log_2
    n$.  Hence, at the end of Phase I, the number of healthy persons
    remaining on the watch list is, with high probability, $\leq 5k$.
    Cf.\ \cite[(19)]{PrS20}.

    % final filter
    Next we analyze Phase II.  Note that each person has a probability
    of $1/16$ to share a test with an infected, and note that each
    person participates in $\log_2 k$ tests in Phase II.  We deduce the
    probability that a healthy person is unlucky, always sharing a test
    with someone infected, and staying suspicious is $1/16^{\log_2 k} =
    k^{-4}$.  The probability that at least one healthy person is
    unlucky is at most $5k \cdot k^{-4} = 5k^{-3}$.  This is the
    probability of the WA event.

\subsection{TLE analysis: how many have been on the watch list?}

    % time complexity as a random variable
    In this subsection, we study the complexity of the decoder.  Since
    the complexity is linear in the number of prefixes the decoder
    handled, it suffices to study the sum of $|\Wg|$ over various
    lengths $\ell$.

    % formal p.g.f.
    Now, following the same strategy as \eqref{for:prod}, one sees that
    the probability-generating function of the number of prefixes that
    have length $\leq \ell$ and have ever been on $\Wg$ is
    \[ H_\ell(q) = \prod_{i=0}^{\ell-\log_2k} h_i (q) ^k, \]
    where $h_i$ is the probability-generating function of $M_i$; $M_i$
    is the number of descendants one innocent prefix can reproduce in
    $i$ generations.

    % relax to total progeny
    Since $M_i$ increases in $i$ and we want an upper bound, we can
    instead study the limit of $M_i$:
    \[
        M_\infty \coloneqq \lim_{i\to\infty} M_i
        \qquad\text{and}\qquad
        h_\infty \coloneqq \lim_{i\to\infty} h_i.
    \]
    $M_\infty$ is called the \emph{total progeny}, and $h_\infty$ is its
    probability-generating function.  We borrow a result of Dwass.

    \begin{lemma} [\cite{Dwa69}]
        Let $M_\infty$ be the total progeny of some branching process
        whose number of children is controlled by the
        probability-generating function $f(t)$, then
        \[ \Prob \{M_\infty{=}m\} = \frac1m [t^{m-1}] f(t)^m, \]
        i.e., the $(m - 1)$th coefficient of $f(t)^m$ divided by $m$.
    \end{lemma}

    % p.g.f. for total progeny \simple{Did that.}
    In our case, $f(t)$ is $(15 + t^2)/16$.  The binomial theorem
    dictates that $f(t)^m \coloneqq \sum_{i=0}^m \binom{m}{i}
    (15/16)^{m-i} (1/16)^i t^{2i}$.  So
    \begin{align*}
        h_\infty(q)
        & = \sum_{m=1}^\infty \frac{q^m}{m} \cdot [t^{m-1}] f(t)^m \\
        & = \sum_{j=0}^\infty \frac{q^{2j+1}}{2j+1} 
            \binom{2j+1}{j} \frac{15^{j+1}}{16^{2j+1}}.
        \shortintertext{
            These are the Catalan numbers \cite[OEIS: A000108]{OEIS}:
        }
        h_\infty(q)
        & = \frac{15q}{16} \sum_{j=0}^\infty \frac{1}{j+1} 
            \binom{2j}{j} \Bigl( \frac{15q^2}{16^2} \Bigr)^j \\
        & = \frac{15q}{16} \cdot \frac
            {1 - \sqrt{1 - 4 \cdot 15q^2/16^2}}
            {2 \cdot 15q^2/16^2}.
    \end{align*}
    The generating function converges absolutely whenever the
    sub-formula inside the square root is positive.

    % apply Chernoff bound
    Now that we obtain $h_\infty(2) = 3$,
    \[ H_\ell(2) \leq \prod_{i=0}^\ell h_\infty (2) ^k = 3^{k\ell}. \]
    Invoke the Chernoff bound to control the number of prefixes ever
    placed on $\Wg$:
    \begin{align*}
        \kern2em
        & \kern-2em
        \Prob \{ \text{number of prefixes handled} \geq 3k\log_2n \} \\
        & \leq \frac {H_{\log_2n} (2)} {2^{3k\log_2n}}
        \leq \frac {\exp(\ln(3) k \log_2 n)} {\exp(3 \ln(2) k \log_2 n)}
        \leq n^{-k}.
    \end{align*}
    Cf.\ \cite[(13)]{PrS20}.  This suggests that the time complexity is,
    with high probability, linear in $k \ln n$.  Therefore, if we set a
    timer of $O(k \ln n)$ and terminate the decoder after the alarm
    sounds, we increase the error probability by a negligible amount,
    but constrain the time complexity to that of the best possible
    order.

    % average total progeny
    A remark on the average behavior is that, as $h_\infty'(1) = 8/7$,
    the number prefixes ever handled is a random variable whose average
    is at most $(1 + 8/7) \* k \* \log_2(n/k)$.  Cf.\ \cite[(1)]{PrS20}.

\subsection{Section wrap up}

    In this section, we assume that the $k$ prefixes of length $\log_2
    k$ each includes an infected person.  If this is not the case, the
    recurrence formula \eqref{for:q^k} will not hold.  However, this
    only means that $q^k$ in \eqref{for:q^k} will be replaced by a
    smaller power, so all the upper bounds given in this section still
    hold.  This concludes PCNS's main theorem.

    \begin{theorem} [\cite{ChN20, PrS20}]
        Fix a $\theta \in (0, 1)$.  For $n$ large enough,
        there exists a measurement matrix that
        (a) performs $m = 16 \* k \* \log_2 n$ tests,
        (b) pairs with a decoder with $O(k \* \ln n)$ operations,
        (c) produces no false negatives, and
        (d) produces one or more false positives with probability at
            most $5 k^{-3} + e^{-k} + n^{-k}$.
    \end{theorem}

\section{Our Change to Prove Theorem~\ref{thm:COMP}}    \label{sec:COMP}

    In this section, we modify PCNS to prove Theorem~\ref{thm:COMP}.
    The key is to mod $ck$ instead of modding $16k$, where $c \coloneqq
    1 / \ln(2 - 4\ee)$ for some very small number $\ee > 0$.  Choosing
    this factor will make us stay as close to COMP's code rate as
    possible while keeping the attracting behavior of the list size.

\subsection{Phase I of PCNS--COMP}                    \label{sec:COMP-I}

    % COMP I testing schedule
    Assuming $ck$ is an integer, we parameterize the tests in Phase I by
    $\{\log_2(k) + 1, \dotsc, \log_2 n\} \times \{0, \dotsc, ck - 1\}$.
    Assign persons with prefix $P \in \{0, 1\}^\ell$, for each $\ell$,
    to the test $(\ell, \Hash(P) \bmod ck)$.

    % COMP I decoder
    The decoder is the same one as before.  The decoder first
    initializes $\Wp$ with $\{0, 1\}^{\log_2 k}$.  Then, in a loop, it
    puts $P0$ and $P1$ on $\Wg$ for any $P \in \Wp$, empties $\Wp$, puts
    $Q$ on $\Wp$ for if $(\Length(Q), \Hash(Q) \bmod ck)$ is positive,
    and empties $\Wg$.  The loop breaks when the strings in $\Wp$ have
    length $\log_2 n$.

\subsection{Phase II of PCNS--COMP}

    % COMP II testing schedule
    For Phase II, parameterize the tests by $\{1, \dotsc, \log_2 k\}
    \times \{0, \dotsc, ck - 1\}$.  Assign the person with label $S$ to
    the tests $(\ell, \Hash(\ell, S) \bmod ck)$ for all $\ell = 1,
    \dotsc, \log_2 k$.

    % COMP II testing schedule
    To decode, do the following for each and every person $S \in \Wp$:
    check if $(\ell, \Hash(\ell, S) \bmod ck)$ are positive for all $1
    \leq \ell \leq \log_2 k$.  If so, declare that $S$ is infected.

\subsection{WA analysis of PCNS--COMP}

    We are to bound the wrong-answer probability from above.

    % probability of alibi
    Consider an innocent prefix placed in the test labeled $(\ell, 0)$.
    The probability that no infected persons are in this test is
    \[
        \Bigl( 1 - \frac{1}{ck} \Bigr)^k
        = \exp \Bigl( \frac {-(1 + o(1)) k} {ck} \Bigr)
        = \frac {1 + o(1)} {2 - 4\ee}
        > \frac12 + \ee.
    \]
    (The last inequality assumes that $k$ is large enough.) That is to
    say, with probability slightly higher than $1/2$, an innocent prefix
    will be pruned.

    % generating function of children
    With $a \coloneqq 1/2 - \ee$, the probability-generating function
    that represents the number of children becomes $f(q) \coloneqq (1 -
    a) + a q^2$ .  Consider this Möbius transformation that matches
    $f(1)$, $f'(1)$, and $f''(1)$:
    \[
        g(q)
        \coloneqq 1 + \frac{4a(q - 1)}{3 - q}
        = f(q) + \frac {a(q - 1)^3}{3 - q}.
    \]
    $g(q)$ serves as an upper bound on $f(q)$ for $q \in [1, 3]$.
    Similar to \eqref{for:diff}, $(g^{\circ i})'(1) = (2a)^i$.
    Therefore
    \[
        g^{\circ i} (q)
        = 1 + \frac {2 (q - 1)} {(3 - q) (1 + 2\ee)^i + (q - 1)}
    \]
    and, particularly, $g^{\circ i} (2) = 1 + 2 / (1 + (1 + 2\ee)^i)$.

    % generating function of total and Chernoff
    Control the generating function of $N_\ell$:
    \[
        F_\ell(2)
        % \leq \prod_{i=0}^\infty g^{\circ i} (2) ^k
        \leq \exp \Bigl( \sum_{i=0}^\infty
                                      \frac {2} {(1 + 2\ee)^i} \Bigr) ^k
        \leq \exp \Bigl( \frac{1 + 2\ee} {\ee} \Bigr) ^k.
    \]
    Control the tail probability using the Chernoff bound:
    \[
        \Prob \{N_\ell{\geq}2k/\ee\}
        \leq \frac{F_\ell(2)}{2^{2k/\ee}}
        \leq e^{-k}. \]
    The last inequality needs $\ee < 1/8$.  This proves that the number
    of healthy persons remaining on the watch list at the end of Phase I
    is very likely $\leq 2k/\ee$.

    % filter
    Now for Phase II, each person has a probability of $a = 1/2 - \ee$
    to participate in a positive test.  The probability that a healthy
    person is unlucky, always sharing a test with infection, and staying
    suspicious is $(1/2 - \ee)^{\log_2 k} \leq 1 / k^{1+2\ee}$.  By the
    union bound, at least one healthy person being unlucky is rare, only
    $2k/\ee / k^{1+2\ee} = 2 / \ee k^{2\ee}$.  This plus the probability
    of long list ($e^{-k}$) is an upper bound on the WA probability.

\subsection{TLE analysis of PCNS--COMP}

    We are to bound the time-limit-exceeded probability from above.  It
    is clear that $f(t)^m \coloneqq \sum_{i=0}^m \binom mi a^i (1 -
    a)^{m-i} t^{2i}$ and that
    \begin{align*}
        h_\infty(q)
        & = \sum_{m=1}^\infty \frac{q^m}{m} \cdot [t^{m-1}] f(t)^m \\
        & = \sum_{j=0}^\infty \frac{q^{2j+1}}{2j + 1} 
            \binom{2j + 1}{j} a^j (1 - a)^{j+1} \\
        & = (1 - a) q \cdot \sum_{j=0}^\infty \frac{1}{j+1} 
            \binom{2j}{j} (a (1 - a) q^2)^j \\
        \shortintertext{(by the generating function of Catalan numbers)}
        & = (1 - a) q \cdot \frac
            {1 - \sqrt{1 - 4 \cdot a (1 - a) q^2}}
            {2 \cdot a (1 - a) q^2}
    \end{align*}
    Verifying $a (1 - a) (1 + \ee^2)^2 \leq 1/4$, one deduces
    $h_\infty(1 + \ee^2) \leq (1 - a) q \cdot 2 < 2$ (assuming $\ee <
    1/3$).  One then bounds the tail probability:
    \begin{align*}
        \kern2em
        & \kern-2em
        \Prob \{ \text{number of hashing} \geq \ee^{-2} \log_2 n \} \\
        & \leq \frac {H_{\log_2n} (1 + \ee^2)}
            {\exp( \ln(1 + \ee^2) \ee^{-2} k \log_2 n)}
        \leq 2 n^{-k}.
    \end{align*}
    (Using $\ee < 1$.) This is an upper bound on the TLE probability.

\subsection{Section wrap up}

    We described a modified GT scheme based on Price--Scarlett
    \cite{PrS20} and Cheraghchi--Nakos \cite{ChN20}.
    We arranged $\ln(2 - 4\ee)^{-2} k \log_2 n$ tests.
    We saw that the probability of the WA event is $2 / \ee k^{2\ee} +
    e^{-k}$ and the only type of error is false positives.  We showed
    that the time complexity is $O(\ee^{-2} k \ln n)$ with TLE
    probability $2 n^{-k}$.  Together, they prove
    Theorem~\ref{thm:COMP}.

\section{Our Change to Prove Theorem~\ref{thm:DD}}        \label{sec:DD}

    In this section, we modify PCNS further to prove
    Theorem~\ref{thm:DD}.  The new idea introduced here is, instead of
    baking the prefixes all the way up to length $\log_2 n$, we stop
    baking at length $\log_2(n/k)$, and call a DD protocol to finish the
    remaining.

\subsection{Phase I of PCNS--DD}

    % DD testing schedule
    The strategy for Theorem~\ref{thm:DD} consists of two phases.  In
    Phase I, tests are parameterized by $\{\log_2(k) + 1,\allowbreak
    \dotsc, \log_2(n/k)\} \times \{0, \dotsc, ck - 1\}$, where $c
    \coloneqq 1 / \ln(2 - 4\ee)$.  Persons with prefix $P \in \{0,
    1\}^\ell$ are assigned to the test $(\ell, \Hash(P) \bmod ck)$.

    % DD decoder
    The decoder will enter the exact same loop as in
    Section~\ref{sec:COMP-I}, except that the loop will break when the
    length reaches $\log_2(n/k)$.  Accompanied by the same analysis, we
    expect $2k/\ee$ prefixes remaining on the watch list.  Each prefix,
    judging by its length, includes $k$ persons.

\subsection{Phase II of PCNS--DD}

    % want (adaptive) DD
    For Phase II, we attempt to perform an optimal DD scheme on those
    $2k/\ee \cdot k = 2k^2/\ee$ suspects.  We do so by assigning persons
    to random tests with help of the hash function.

    % actual declaration of testing schedule
    Let there be tests indexed by $\{0, ..., ck \log_2(2k/\ee) - 1\}$.
    For each person $S \in \{0, 1\}^{\log_2 n}$, assign $S$ to
    $\Hash(\ell, S) \bmod ck \log_2(2k/\ee)$ for all integers $\ell \in
    [1, c \ln(2k/\ee)]$.

    % decoding
    To decode Phase II, it is enough to focus on ones whose prefixes
    remain on $\Wp$.  There are, with high probability, $\leq 2k^2/\ee$
    such persons.  So, effectively, we are running the DD algorithm on
    $N \coloneqq 2K^2/\ee$ individuals, $K \coloneqq k$ infected
    individuals, $T \coloneqq cK \log_2(N/K)$ tests, and column weight
    $L \coloneqq c \ln(N/K)$.  But this is the exact situation faced by
    \cite[Section IV-F]{JAS19} (proof of Theorem~3 therein).  Hence we
    know, by the cited theorem, that DD succeeds with probabilities
    converging to one.

    % complexity
    As for the complexity, recall that DD consists of three steps.  Step
    One: construct a bipartite graph; one part for persons, the other
    part for tests.  Step Two: delete negative tests and all persons
    participating in those.  Step Three: find the test with degree one
    and declare that the only person it connects to is infected.
    Clearly the complexity of Step One is the number of persons times
    the number of tests each person is in; it is $2k^2/\ee \cdot c
    \ln(2k/\ee)$.  The complexity of Step Two is the number of edges,
    which is also $2k^2/\ee \cdot c \ln(2k/\ee)$.  The complexity of
    Step Three is the number of tests.  Overall, the complexity is
    $O(\ee^{-1} k^2 \ln(2k/\ee))$.

\subsection{Section wrap up}

    $ck \log_2(n/k^2)$ tests are in Phase I and $ck \log_2(2k/\ee)$ in
    Phase II.  That is $ck \log_2(2n/\ee k)$ in total.  The WA
    probability is the probability that the list size at the end of
    Phase I exceeds $2k/\ee$, which is $e^{-k}$, plus the probability of
    false negatives of DD, which is implicit but close to $0$.  The
    complexity is $O(\ee^{-2} k \ln(n/k))$ (with TLE probability
    $(n/k)^{-k} < k^{-k}$) for Phase I and $O(\ee^{-1} k^2 \ln(2k/\ee))$
    for Phase II.  Together, they prove Theorem~\ref{thm:DD}.

\section{Acknowledgment}

    This work was supported by NSF grants CCF-2107346 and CCF-2210823
    and a Simons Investigator award.

\tracingpages1
\IEEEtriggeratref{10}
% \IEEEtriggercmd{\enlargethispage{\dimexpr 390pt -518.2922pt}\pagebreak}
\bibliographystyle{IEEEtran}
\bibliography{Bonsai-31.bib}

% Generated by IEEEtran.bst, version: 1.14 (2015/08/26)
\begin{thebibliography}{10}
\providecommand{\url}[1]{#1}
\csname url@samestyle\endcsname
\providecommand{\newblock}{\relax}
\providecommand{\bibinfo}[2]{#2}
\providecommand{\BIBentrySTDinterwordspacing}{\spaceskip=0pt\relax}
\providecommand{\BIBentryALTinterwordstretchfactor}{4}
\providecommand{\BIBentryALTinterwordspacing}{\spaceskip=\fontdimen2\font plus
\BIBentryALTinterwordstretchfactor\fontdimen3\font minus \fontdimen4\font\relax}
\providecommand{\BIBforeignlanguage}[2]{{%
\expandafter\ifx\csname l@#1\endcsname\relax
\typeout{** WARNING: IEEEtran.bst: No hyphenation pattern has been}%
\typeout{** loaded for the language `#1'. Using the pattern for}%
\typeout{** the default language instead.}%
\else
\language=\csname l@#1\endcsname
\fi
#2}}
\providecommand{\BIBdecl}{\relax}
\BIBdecl

\bibitem{ChN20}
M.~Cheraghchi and V.~Nakos, ``Combinatorial group testing and sparse recovery schemes with near-optimal decoding time,'' in \emph{2020 IEEE 61st Annual Symposium on Foundations of Computer Science (FOCS)}, Nov 2020, pp. 1203--1213.

\bibitem{PrS20}
E.~Price and J.~Scarlett, ``{A Fast Binary Splitting Approach to Non-Adaptive Group Testing},'' in \emph{Approximation, Randomization, and Combinatorial Optimization. Algorithms and Techniques (APPROX/RANDOM 2020)}, vol. 176, 2020, pp. 13:1--13:20.

\bibitem{AlE22}
M.~Aldridge and D.~Ellis, \emph{Pooled Testing and Its Applications in the COVID-19 Pandemic}.\hskip 1em plus 0.5em minus 0.4em\relax Cham: Springer International Publishing, 2022, pp. 217--249.

\bibitem{DuH93}
D.-Z. Du and F.~K. Hwang, \emph{Combinatorial Group Testing and Its Applications}.\hskip 1em plus 0.5em minus 0.4em\relax World Scientific, 1993.

\bibitem{AJS19}
M.~Aldridge, O.~Johnson, and J.~Scarlett, ``Group testing: An information theory perspective,'' \emph{Foundations and Trends{\textregistered} in Communications and Information Theory}, vol.~15, no. 3-4, pp. 196--392, 2019.

\bibitem{AtS09}
G.~Atia and V.~Saligrama, ``Noisy group testing: An information theoretic perspective,'' in \emph{2009 47th Annual Allerton Conference on Communication, Control, and Computing (Allerton)}, Sep. 2009, pp. 355--362.

\bibitem{ScC17}
J.~Scarlett and V.~Cevher, ``Limits on support recovery with probabilistic models: An information-theoretic framework,'' \emph{IEEE Transactions on Information Theory}, vol.~63, no.~1, pp. 593--620, Jan 2017.

\bibitem{CGH20}
A.~Coja-Oghlan, O.~Gebhard, M.~Hahn-Klimroth, and P.~Loick, ``Information-theoretic and algorithmic thresholds for group testing,'' \emph{IEEE Transactions on Information Theory}, vol.~66, no.~12, pp. 7911--7928, Dec 2020.

\bibitem{INR10}
P.~Indyk, H.~Q. Ngo, and A.~Rudra, ``Efficiently decodable non-adaptive group testing,'' in \emph{Proceedings of the 2010 Annual ACM-SIAM Symposium on Discrete Algorithms (SODA)}, 2010, pp. 1126--1142.

\bibitem{IKW19}
H.~A. Inan, P.~Kairouz, M.~Wootters, and A.~Ozgur, ``On the optimality of the kautz-singleton construction in probabilistic group testing,'' in \emph{2018 56th Annual Allerton Conference on Communication, Control, and Computing (Allerton)}, Oct 2018, pp. 188--195.

\bibitem{CGH21}
A.~Coja-Oghlan, O.~Gebhard, M.~Hahn-Klimroth, and P.~Loick, ``Optimal group testing,'' \emph{Combinatorics, Probability and Computing}, vol.~30, no.~6, pp. 811--848, 2021.

\bibitem{LCP19}
K.~Lee, K.~Chandrasekher, R.~Pedarsani, and K.~Ramchandran, ``Saffron: A fast, efficient, and robust framework for group testing based on sparse-graph codes,'' \emph{IEEE Transactions on Signal Processing}, vol.~67, no.~17, pp. 4649--4664, Sep. 2019.

\bibitem{BCS21}
S.~Bondorf, B.~Chen, J.~Scarlett, H.~Yu, and Y.~Zhao, ``Sublinear-time non-adaptive group testing with o(k log n) tests via bit-mixing coding,'' \emph{IEEE Transactions on Information Theory}, vol.~67, no.~3, pp. 1559--1570, March 2021.

\bibitem{BJA13}
L.~Baldassini, O.~Johnson, and M.~Aldridge, ``The capacity of adaptive group testing,'' in \emph{2013 IEEE International Symposium on Information Theory}, July 2013, pp. 2676--2680.

\bibitem{Ald17}
M.~Aldridge, ``The capacity of bernoulli nonadaptive group testing,'' \emph{IEEE Transactions on Information Theory}, vol.~63, no.~11, pp. 7142--7148, Nov 2017.

\bibitem{wiki-gt}
{Wikipedia contributors}, ``Group testing --- {Wikipedia}{,} the free encyclopedia,'' \url{https://en.wikipedia.org/w/index.php?title=Group_testing&oldid=1131369566}, 2023, [Online; accessed 9-January-2023].

\bibitem{JAS19}
O.~Johnson, M.~Aldridge, and J.~Scarlett, ``Performance of group testing algorithms with near-constant tests per item,'' \emph{IEEE Transactions on Information Theory}, vol.~65, no.~2, pp. 707--723, Feb 2019.

\bibitem{BSP22}
W.~H. Bay, J.~Scarlett, and E.~Price, ``{Optimal non-adaptive probabilistic group testing in general sparsity regimes},'' \emph{Information and Inference: A Journal of the IMA}, vol.~11, no.~3, pp. 1037--1053, 02 2022.

\bibitem{FlM21}
L.~Flodin and A.~Mazumdar, ``Probabilistic group testing with a linear number of tests,'' in \emph{2021 IEEE International Symposium on Information Theory (ISIT)}, July 2021, pp. 1248--1253.

\bibitem{ABJ14}
M.~Aldridge, L.~Baldassini, and O.~Johnson, ``Group testing algorithms: Bounds and simulations,'' \emph{IEEE Transactions on Information Theory}, vol.~60, no.~6, pp. 3671--3687, June 2014.

\bibitem{CJB17}
S.~Cai, M.~Jahangoshahi, M.~Bakshi, and S.~Jaggi, ``Efficient algorithms for noisy group testing,'' \emph{IEEE Transactions on Information Theory}, vol.~63, no.~4, pp. 2113--2136, April 2017.

\bibitem{ChN20x}
M.~Cheraghchi and V.~Nakos, ``Combinatorial group testing and sparse recovery schemes with near-optimal decoding time,'' 2020.

\bibitem{Dwa69}
M.~Dwass, ``The total progeny in a branching process and a related random walk,'' \emph{Journal of Applied Probability}, vol.~6, no.~3, pp. 682--686, 1969.

\bibitem{OEIS}
{OEIS Foundation Inc.}, ``The {O}n-{L}ine {E}ncyclopedia of {I}nteger {S}equences,'' 2023, published electronically at \url{http://oeis.org}.

\end{thebibliography}

\end{document}